\documentclass[]{spie}  

\usepackage{amsmath,amsfonts,amssymb}
\usepackage{graphicx}
\usepackage[colorlinks=true, allcolors=blue]{hyperref}

\newcommand{\be}{\begin{equation}}
\newcommand{\ee}{\end{equation}}

\title{Superkicks and momentum density tests via micromanipulation
}

\author[a]{Andrei Afanasev}

\affil[a]{Department of Physics,
The George Washington University, Washington, DC 20052, USA}

\author[b]{Carl E. Carlson}

\affil[b]{Physics Department, William \& Mary, Williamsburg, Virginia 23187, USA}

\author[c]{Asmita Mukherjee}

\affil[c]{Department of Physics, Indian Institute of Technology Bombay, Powai, Mumbai 400076, India}

\authorinfo{Further author information: (Send correspondence to C.C.)\\C.C.: E-mail: carlson@physics.wm.edu}

\begin{document} 
\maketitle

\begin{abstract}
There is an unsettled problem in choosing the correct expressions for the local momentum density and angular momentum density of electromagnetic fields (or indeed, of any non-scalar field).  
If one only examines plane waves, the problem is moot, as the known possible expressions all give the same result. 
The momentum and angular momentum density expressions are generally obtained from the energy-momentum tensor, in turn obtained from a Lagrangian.
The electrodynamic expressions obtained by the canonical procedure are not the same as the symmetric Belinfante reworking. 
For the interaction of matter with structured light, for example, twisted photons, this is important; there are drastically different predictions for forces and angular momenta induced on small test objects.  
We show situations where the two predictions can be checked, with numerical estimates of the size of the effects.
\end{abstract}

\keywords{Structured light, Twisted photons, Belinfante tensor}

\section{INTRODUCTION}
\label{sec:intro}  

Twisted light is a light beam with intrinsic orbital angular momentum (OAM) in the direction of motion; for reviews, see~\cite{2011AdOP....3..161Y,Bliokh:2015doa}.   The components of the vector potential that are transverse to the overall propagation direction of the beam can be given by~\cite{2019PhRvA..99b3845Q}
\be
\vec A_\perp(\vec r, t) =  A_0 \hat \epsilon_\Lambda \,  e^{i(k_z z-\omega t)}\,  e^{i \ell \phi} \,J_\ell(\kappa \rho)   \,.
\ee
$A_0$ is a normalization constant, $\hat\epsilon$ is a polarization vector which can be made more specific as  $\hat\epsilon_\Lambda$ for circular polarization $\Lambda = \pm 1$.  In wave number space, or momentum space quantum mechanically, it is a superposition of plane wave states all with the same longitudinal momentum $k_z$ and the same magnitude transverse momentum $\kappa$ but varying azimuthal directions.  The phase $e^{i \ell \phi}$ makes the orbital angular momentum in the direction of motion be $\ell$ (or $\ell \hbar$), and the total angular momentum is $m_\gamma = \ell + \Lambda$.
The swirling wavefront, in addition to longitudinal momentum, has transverse momentum components that can twist objects in its path or give them sideways kicks.  

The expressions that give the momentum and angular momentum densities are obtained from the energy-momentum tensor~\cite{Jauch:1976ava,Bjorken:1965zz}.
Using the electromagnetic Lagrangian 
$L = - 
F_{\alpha\beta} F^{\alpha\beta}/(4\mu_0)$ and the canonical or Noether procedure gives the canonical energy-momentum tensor
\begin{equation}
    T^{\mu\nu} =	- \frac{1}{\mu_0} F^{\mu\alpha} \partial^\nu \! A_\alpha - g^{\mu\nu} L    \,.
\end{equation}
It is not symmetric in its two indices.  Aside from aesthetics, the field equation in General Relativity requires a symmetric energy-momentum tensor.  It can be symmetrized by adding a total derivative~\cite{1940Phy.....7..449B,rosenfeld1940energy}, whereby spatial integrals over the energy-momentum tensor remain unchanged.  Doing so gives the Belinfante tensor
\begin{equation}
    \theta^{\mu\nu} = - \frac{1}{\mu_0} F^{\mu\alpha} F^\nu_{\ \alpha} -	g^{\mu\nu} L    \,.
\end{equation}
The momentum densities $\vec {\mathcal P}$ are obtained from $T^{0i}\!\,/c$ or $\theta^{0i}\!\,/c$,  and are
\begin{equation}
    \vec {\mathcal P} = 
			 \epsilon_0 \vec E \cdot (\vec\nabla) \vec A	\,,		\quad \text{canonical}\ 
\end{equation}
or  
\begin{equation}
    \vec {\mathcal P} = 
			 \epsilon_0 \vec E \times \vec B		\,,		\quad \ 	\text{Belinfante}.
\end{equation}
In general, they are numerically distinct.
The related results for angular momentum density $\vec {\mathcal J}$ are
\begin{align}
\vec {\mathcal J} = \left\{
			\begin{array}{l}
	\epsilon_0 \vec E \cdot (\vec r \times \vec\nabla) \vec A	+ \epsilon_0 \vec E \times \vec A
		\,,		\quad\, \text{canonical},	\\
			 \epsilon_0 \, \vec r \times ( \vec E \times \vec B )		\,,		
		\qquad\qquad\qquad	\text{Belinfante}.		
			\end{array}
\right.
\end{align}
Again, the two expressions differ by just a total derivative.  But they do differ locally, so do not lead to the same torque upon small test objects.

Interestingly, an alternative way to obtain the momentum density of electromagnetic field is to infer it from forces on small dielectrics calculated from Lorentz force law~\cite{2009PhRvL.102k3602A}.  The momentum density obtained this way agrees with the canonical result.  Further discussions of momentum density   definitions can be found in~\cite{Bliokh:2015doa,2013EJPh...34.1337B,1978OptCo..24..185H,2019PhRvA..99f3832W,2014OExpr..22.6586O,2002PhRvL..88e3601O,2003PhRvL..91i3602G}.

\section{Angular and transverse momentum density tests}

We will discuss some predictions for making objects spin by shining twisted light upon them, thinking in one case in terms of the torque generated by absorbing the electromagnetic angular momentum, and in another case in terms of the sideways kick, sometimes called the superkick~\cite{2013JOpt...15l5701B}, given to individual particles in a system.

\begin{figure} [b]
   \begin{center}
   \begin{tabular}{c} 
   \includegraphics[height=5cm]{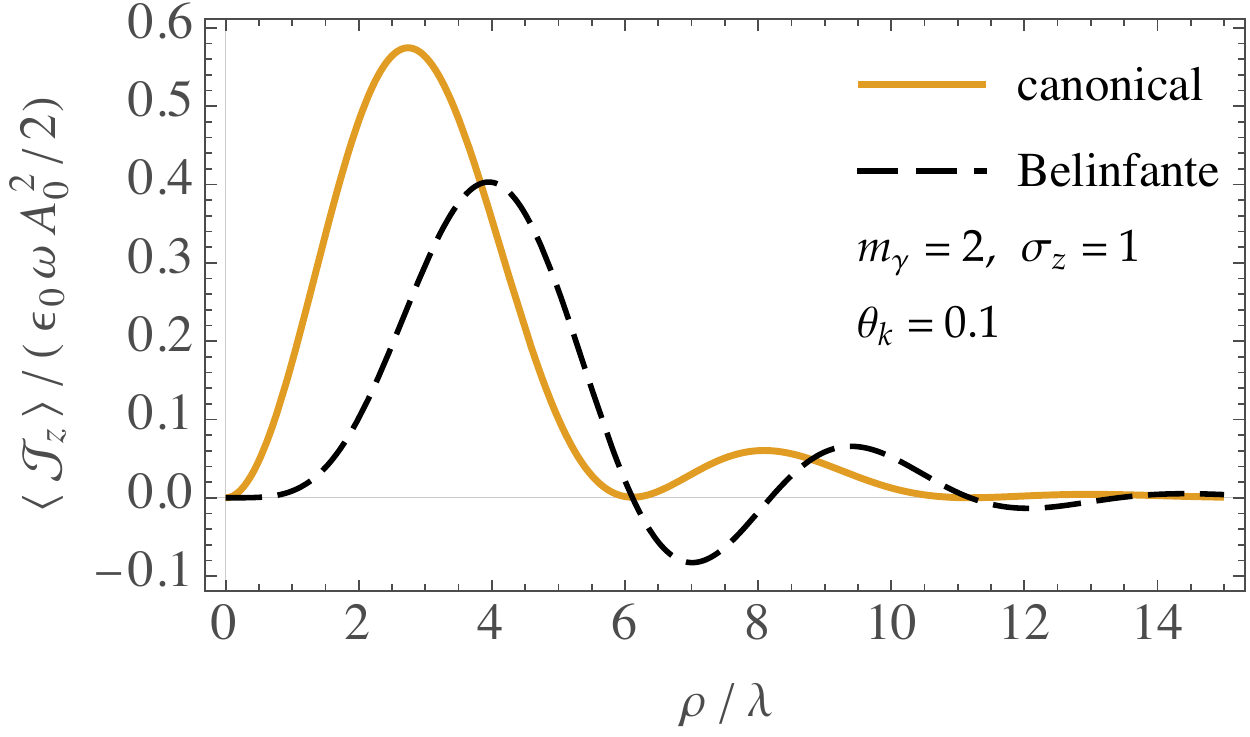}
   \end{tabular}
   \end{center}
   \caption[example] 
   { \label{fig:example} 
Angular momentum density on a ring of radius $\rho$ for a twisted light beam of total angular momentum $m_\gamma =2$  and circular polarization $\sigma_z =1$, with angular frequency $\omega$ and $A_0$ normalizing the strength of the beam's electric field.  The pitch angle $\theta_k$ is $\arctan(\kappa/k_z)$.
The Belinfante case has regions where the angular momentum density swirls in a direction opposite to the overall angular momentum.}
   \end{figure}

One specific result for the $z$-component (averaged over time) of the angular momentum density, plotted versus radius or the distance $\rho$ out from the vortex axis, is shown in Fig.~\ref{fig:example}.  In this figure, the projected total angular momentum of the field is $m_\gamma=2$ and the circular polarization is $\Lambda=\sigma_z = 1$.  One sees significant  differences between the canonical and Belinfante results at most radii.


\paragraph{One test:  twisted light on a cylindrical shell.}

Given the plot just mentioned, we will consider shining a twisted beam on a cylindrical shell, or hollow cylinder.  (Note that the torque on a filled cylinder would be an integrated torque, which is the about the same for the two cases, depending on the radius of the cylinder.)

\begin{figure} [ht]
   \begin{center}
   \begin{tabular}{c} 
   \includegraphics[height=4.5cm]{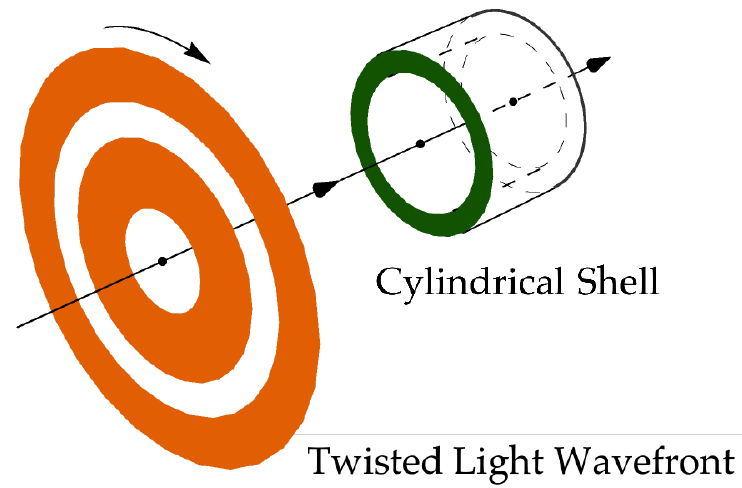}
	\end{tabular}
	\end{center}
   \caption[example] 
   { \label{fig:hollow} 
Twisted light hitting a hollow cylinder, with axes coincident.}
   \end{figure}

Let the cylinder and beam axes be coincident, Fig.~\ref{fig:hollow} and the torque on shell will be proportional to angular momentum absorbed from beam at radius of cylinder.


For radius $\rho = 2 \,\mu\text{m} \approx 2.74 \,\lambda$ and other selected parameters, and cylindrical shell sitting in a kerosine bath (the viscous drag is calculable~\cite{1995PhRvL..75..826H,1996PhRvA..54.1593F}), terminal rotation frequency is
\begin{equation}
    f  \approx
0.55   \text{\,Hz} ,	
\quad \text{canonical} , 
\end{equation}
or
\begin{equation}
    f  \approx  
0.23   \text{\,Hz} ,	
\quad \text{Belinfante}.
\end{equation}
The other parameters used in this example are  $\lambda = 0.729 \,\mu$m, power in beam = 4 mW, the beam put inside a Gaussian envelope of width 10 $\lambda$,  thickness of shell = 0.5 $\mu$m,  length of cylinder = 2  $\mu$m, density of shell material twice water.

\paragraph{Another test: a two-ion rotor.}

We will study this case in terms of the effect of transverse momentum kicks, or superkicks, upon atoms constrained to move in a circle.  

Consider a rotor has two $^{40}$Ca ions constrained to revolve in a circular path in a plane.  The atoms are well constrained to stay in the plane, and their radius from the center of the circle is well fixed, but they are free to revolve in a circle about the center.  Such a rotor exists, and is described in~\cite{2019PhRvL.123m3202U}, but has not been used for the present purpose.  


   \begin{figure} [ht]
   \begin{center}
   \begin{tabular}{c} 
   \includegraphics[height=4cm]{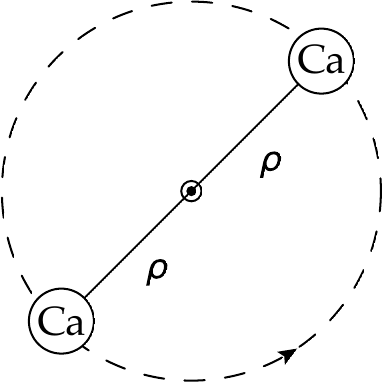}
	\end{tabular}
	\end{center}
   \caption[example] 
   { \label{fig:carotor} 
A two-ion Calcium rotor.}
   \end{figure} 
   
\begin{figure} [t]
   \begin{center}
   \begin{tabular}{c} 
   \includegraphics[height=5cm]{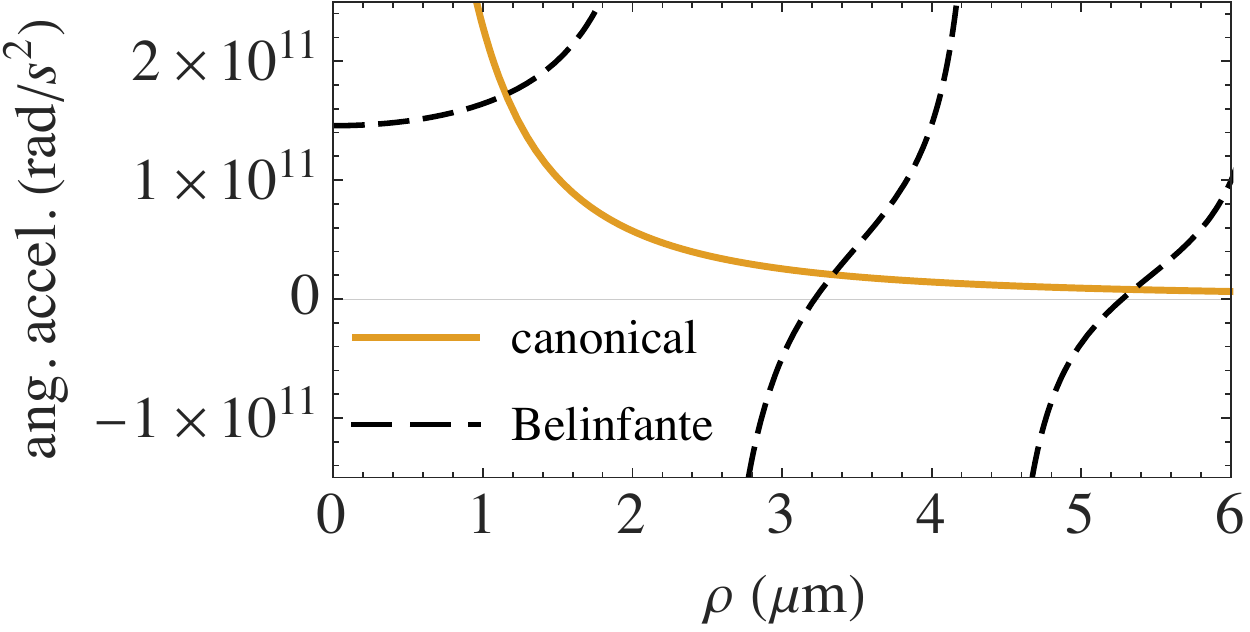}
	\end{tabular}
	\end{center}
   \caption[example] 
   { \label{fig:accel} 
Calculated angular acceleration for a two-ion Calcium rotor of varying radii, with further description in the text.}
   \end{figure}

Shine the beam perpendicular to plane of the rotor with vortex line passing thru its center. The beam can excite either ground state ion, and the ion receives one transverse momentum kick per lifetime of the excited state.  This will give a force $dp_\perp/dt$,  and a torque, and an angular acceleration.
The transverse momentum kick per absorption depends on the ratio of the momentum density and photon number density, and the numerator and denominator of this ratio give a simple result for the canonical case,
\begin{equation}
    p_{\perp} = 
		 \frac{  \ell \, \hbar} {\rho}
		\qquad \qquad \quad \, \text{canonical},
\end{equation}
and
\begin{equation}
    p_{\perp} = 
			\hbar \kappa 
			\displaystyle  \frac{ J_{\ell+\Lambda}(\kappa \rho) }{ J_{\ell}(\kappa \rho) }
			 \qquad \text{Belinfante}	.
\end{equation}
Again, $\Lambda$ is the photon helicity; a plot of the anticipated angular acceleration for $\Lambda = 1$ and $\ell = 1$ is shown in Fig.~\ref{fig:accel}.
The plot was made using the $4s_{1/2}  \to 4p_{3/2}$ or $4p_{1/2}$ transitions, where the excited state is not metastable but has a fast spontaneous
decay. The situation is analogous to laser cooling~\cite{1975OptCo..13...68H}:
the spontaneous decay is isotropic so statistically there
is no momentum kick in the decay, but the excitation
always involves a momentum kick in the same azimuthal
direction. 
The predicted angular accelerations are wildly different, as Fig.~\ref{fig:accel} shows.


\section{Closing remarks}
\label{sec:closing}


More details and additional situations contrasting the canonical and Belinfante momentum and angular momentum density expressions can be found in~\cite{Afanasev:2022vgl}.  

The canonical and the Belinfante versions of the electromagnetic energy-momentum tensor are by design the same for integrated quantities such as the total momentum or total angular momentum of the field.  However,  point by point in space they are different, and on small test objects they give different results in  calculations of the force or torque from electromagnetic waves striking them.  The differences are only apparent if the light has a structured wave front, and we have worked out examples using the particular example of twisted photons.  In certain regions the differences are especially dramatic.  Not discussed here are tractor beam effects in the Belinfante case, discussed in~\cite{Afanasev:2022vgl} and also noticed in~\cite{Novitsky:07}.  However, the dramatic tractor beam effects lie in limited spatial regions and are sensitive to details of the beam preparation. On the other hand, the dramatic torque differences which we have focused on in this talk are robust and exist over broad spatial regions and could well be confirmed or denied experimentally using ringlike or end-weighted rotor 

\section*{Addendum}

\paragraph{Single particle off-axis in vortex beam.}  At the conference it was also possible to show a third test of the possible angular momentum densities.  This was to put a small test particle off-axis in a twisted photon beam, whence the transverse momentum absorption or the superkick would cause the particle to revolve about the vortex axis.

Such an experiment was actually performed nearly two decades ago~\cite{2003PhRvL..91i3602G}, but measurements were made only at the maxima of the ringlike intensity profile of the twisted beam wavefront.  It happens that the difference between the predictions of the canonical and Belinfante angular momentum is proportional to the derivative of the intensity with respect to distance from the axis~\cite{Afanasev:2022vgl}.  Hence the predictions of the two possibilities are the same at the points where measurements were made.  


   \begin{figure} [t]
   \begin{center}
   \begin{tabular}{c} 
   \includegraphics[height=6cm]{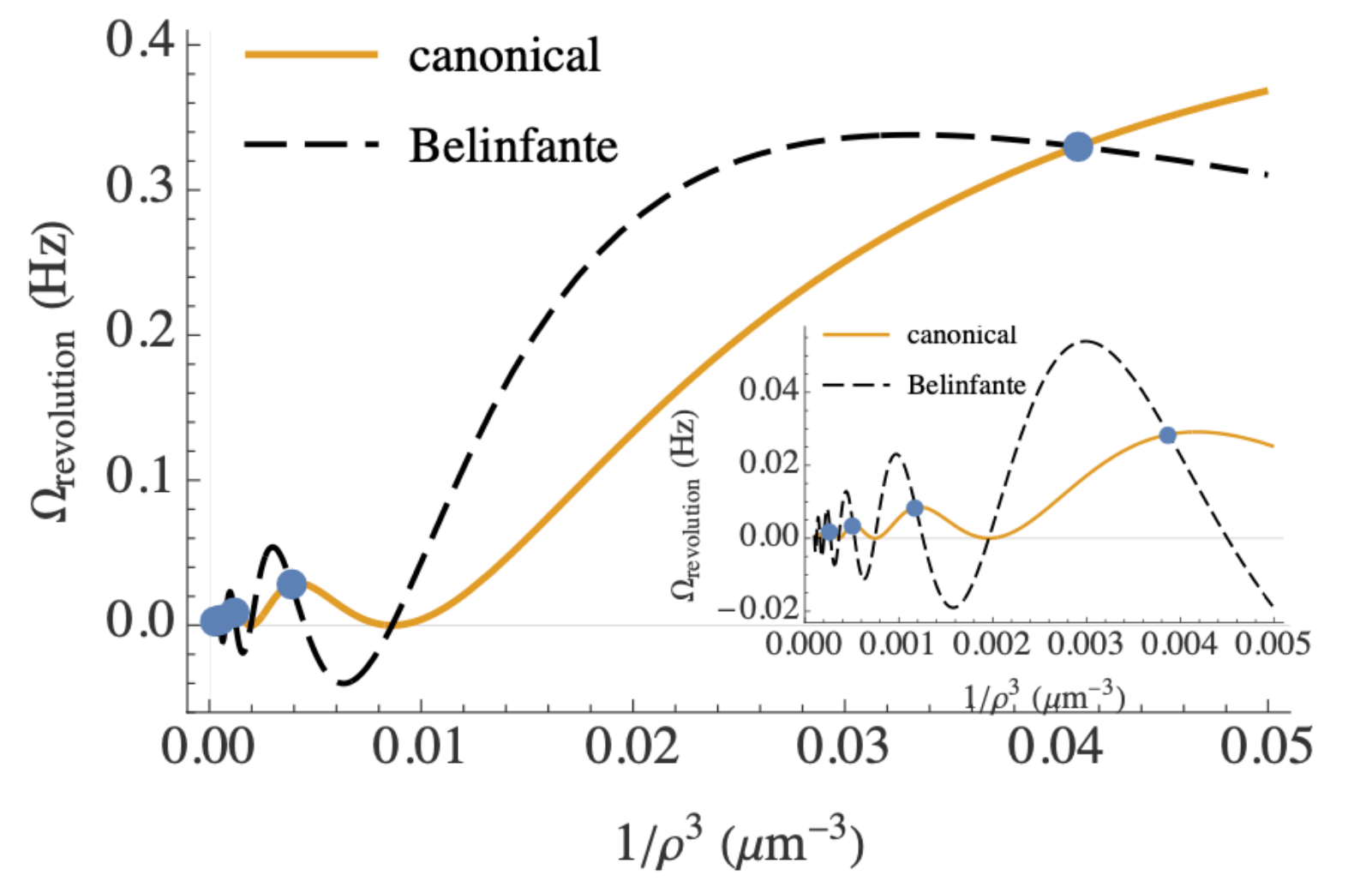}
	\end{tabular}
	\end{center}
   \caption[example] 
   { \label{fig:revolution} 
Calculated angular acceleration for a two-ion Calcium rotor of varying radii, with further description in the text.}
   \end{figure} 
   

Fig.~\ref{fig:revolution} shows the predictions of the Belinfante and canonical angular momentum densities for the off-axis test particle, given as its revolution frequency (for a certain laser power and for $\ell=2$ and $\Lambda=1$) plotted vs.~the distance from the vortex axis cubed, as in~\cite{2003PhRvL..91i3602G}.   The dots show the radii where measurements have been made.  Clearly the predictions are significantly different and measurements with the test particle held at different radii could be quite decisive.  

\acknowledgments 
 
We thank Elliot Leader for stimulating conversations.  A.A. thanks the US Army Research Office Grant W911NF-19-1-0022 for support, C.E.C. thanks the National Science Foundation (USA) for support under grant PHY-1812326, and A. M. thanks the SERB-POWER Fellowship, Department of Science and Technology, Govt. of India  for support.  

\bibliography{momconext} 
\bibliographystyle{spiebib} 

\end{document}